\documentclass[twocolumn]{aastex63}

\usepackage{etoolbox}
\usepackage{graphicx}	
\usepackage{multirow}
\usepackage{cleveref}
\usepackage{CJK}
\usepackage{comment}
\newcommand{\cloudy}{\textsc{cloudy}}
\newcommand{\kms}{km s$^{-1}$}
\newcommand{\cmN}{cm$^{-2}$}

\newcommand{\lam}{$\lambda$}

\newcommand{\lya}{\mbox{Ly$\alpha$}}

\newcommand{\civ}{\mbox{C\,{\sc iv}}}

\newcommand{\siiv}{\mbox{Si\,{\sc iv}}}
\newcommand{\siiii}{\mbox{Si\,{\sc iii}}}

\newcommand{\nv}{\mbox{N\,{\sc v}}}

\newcommand{\ovi}{\mbox{O\,{\sc vi}}}

\newcommand{\pv}{\mbox{P\,{\sc v}}}
\newcommand{\target}{J221531-174408}

\received{}
\revised{}
\accepted{}
\submitjournal{ApJ}
\shorttitle{Line-locking from QSO \target }
\shortauthors{Chen et al.}

\begin{document}

\title{Tracking Outflow using Line-Locking (TOLL). I. The case study of Quasar J221531-174408}

\correspondingauthor{Bo Ma}
\email{mabo8@mail.sysu.edu.cn}

\author{Chen Chen}
\affiliation{Zhuhai College of Science and Technology\\ 
Zhuhai 519000, China}

\author{Weimin Yi}
\affiliation{Yunnan Observatories\\
Kunming 650216, China}

\author{Zhicheng He}
\affiliation{CAS Key Laboratory for Research in Galaxies and Cosmology, Department of Astronomy\\
University of Science and Technology of China\\
Hefei 230026, China}

\author{Fred Hamann}
\affiliation{Department of Physics and Astronomy\\
University of California Riverside\\ 
Riverside, CA 92507, USA}

\author{Bo Ma}
\affiliation{School of Physics $\&$ Astronomy\\
Sun Yat-Sen University\\ 
Zhuhai 519000, China}

\begin{abstract}

Investigating line-locked phenomena within quasars is crucial for understanding the dynamics of quasar outflows, the role of radiation pressure in astrophysical flows, and the star formation history and metallicity of the early universe. 
We have initiated the Tracking Outflow by Line-Locking (TOLL) project to study quasar outflow by studying line-locking signatures using high-resolution high signal-to-noise ratio quasar spectra.
In this paper, we present a case study of the line-locking signatures from QSO \target. 
The spectrum was obtained using the Very Large Telescope-UV Visual Echelle Spectrograph. We first identify associated absorbers in the spectrum using \civ, \nv, and \siiv\ doublets and measure their velocity shifts, covering fractions, and column densities through line profile fitting technique. 
Then we compare the velocity separations between different absorbers, and detect nine pairs of line-locked \civ\ doublets, three pairs of line-locked \nv\ doublets, and one pair of line-locked \siiv\ doublets. 
This is one of the four quasars known to possess line-locked signatures in \civ, \siiv, and \nv\ at the same time.
We also find three complex line-locked systems, where three to five absorbers are locked together through multi-ion doublets. 
Our study suggests that line-locking is a common phenomenon in the quasar outflows, and theoretical models involving more than two clouds and one ionic doublet are needed in the future to explain the formation of these complex line-locking signatures.

\end{abstract}

\section{Introduction}

Quasar outflows play a pivotal role in the evolution of galaxies as they supply kinetic-energy feedback from supermassive black holes to their respective host galaxies \citep{Silk98, Kauffmann00, King03, Scannapieco04, DiMatteo05, Hopkins08, Ostriker10, Debuhr12, Rupke13, Rupke17, Cicone14, Weinberger17, Giustini19, He22, Chen22, Ayubinia23, Naddaf23, Hall24}. 
Line-locking describes a state where the observed velocity difference between distinct kinematic absorption components equals the velocity separation of known atomic transitions. 
This phenomenon, observed within these outflows, provides evidence that radiation pressure significantly influences the dynamics of the clumpy outflow clouds \citep{Goldreich76}.
For example, \citet{Hamann11} reported the detection of multiple narrow absorption line (NAL) systems toward a quasar, which are separated by the velocity separation of the \civ\ doublet \citep[see also][]{Lin20a, Lin20b, Lu20, Chen21, Yi24}. 
Line-locking corresponding to other transitions, such as \ovi\ \citep{Ganguly03, Ganguly13, Mas-Ribas19a, Mas-Ribas19b} and \nv\ \citep{Srianand02, Ganguly03, Mas-Ribas19a, Mas-Ribas19b, Veilleux22, Juranova24}, has also been reported. 
The comprehensive investigation conducted by \cite{Bowler14} revealed that about two-thirds of quasars possessing multiple CIV absorption line systems show characteristics of line-locking, suggesting that such signatures are quite common in quasar outflows. 
However, a more recent examination by \citet{Chen21}, utilizing high-resolution spectral data, indicated that this proportion is as low as 20 to 30 percent. Despite this discrepancy, the consensus remains that line-locking is a prevalent phenomenon.
These findings are of great significance as they aid in understanding of the mechanisms propelling these outflows and their impact on galaxy evolution.

To explain the formation of line-locked signature within active galaxies and the underlying physics, \citet{Lewis23} have studied a simple two-clouds-one doublet model. 
They found that a fine-tuning between the parameters (e.g., column density, ionization parameter) of the two clouds is needed for line-locking to form. 
They emphasized that line-locking serves as a definitive indicator of the dynamical significance of radiation pressure in propelling astrophysical flows. Consequently, investigation these systems provides valuable insights into the role of radiation pressure in the acceleration and dynamics of outflows, which is essential for refining models related to quasar feedback and galaxy formation.

One obstacle to studying line-locking phenomena is that the intermediate widths and often shallow depths of the lines make them difficult to detect (and correctly identify) in spectroscopic surveys using moderate resolution and moderate-to-low signal-to-noise ratios (SNRs), such as those in the Baryon Oscillation Spectroscopic Survey in SDSS-III \citep{Eisenstein11, Dawson13, Paris17}. 
Thus, higher resolution spectra from large ground-based telescopes are crucial for studying line-locked absorption systems from quasars. 
This is indeed can be seen from Fig.3 in \citet{Yi24}, where the shallow \civ\ line-locking signatures are only detected in the higher-resolution HET/LRS-2 spectra. 
We have initiated the Tracking Outflow by Line-Locking (TOLL) project to analyze line-locked absorption systems using high-resolution, high-quality AGN spectra. 
The TOLL project will both analyze detailed properties of line-locked absorbers of individual quasar and conduct statistical distribution of line-locked absorbers for a large sample of quasars. 
In contrast to the study of \citet{Bowler14}, TOLL mainly utilize archived high-resolution, high-quality spectral observations of $\sim$700 quasars from Keck and VLT. 
Some of the spectra are reduced by our group, and the others are taken from \citet{Murphy19} and \citet{OMeara15}. 
The goal of TOLL is to study the quasar outflows from a microscopic point of view, i.e., focus on individual clumpy cloud, and investigate the gaseous environments of quasars.
Here, we present our first case study of the TOLL project, where we simultaneously detected line-locking systems in \civ, \nv, and \siiv\ doublets from the high-resolution spectrum of quasar~\target.

The structure of this paper is outlined as follows: 
Section 2 introduces the UVES spectrum of \target\ and the method employed for fitting the line profiles of associated absorption lines \citep[AALs;][]{Weymann79, Foltz86, Anderson87, Weymann91, Hamann97d} to measure the Doppler parameter, column densities, velocity shifts, and covering fractions of each absorber. 
In Section 3, we present the line-locking pairs detected in this paper and discuss the implications of our results concerning the study of the line-locking phenomenon. 
Section 4 concludes the paper with a summary. Throughout this paper, we utilize a cosmological model characterized by $H_0=71$ \kms\ Mpc $^{-1}$, $\Omega_M=0.27$, and $\Omega_{\Lambda}=0.73$.


\section{QUASAR SPECTRA and DATA ANALYSIS}

\subsection{Quasar Spectra}
The parent sample of quasars for the TOLL project comes from the catalog presented in \citet{chen20}, which are originally obtained with VLT-UVES \citep{Murphy19} and Keck-HIRES \citep{OMeara15}. 
The spectral resolutions are in the range of $22,000\lesssim R\lesssim 71,000$ for VLT-UVES. In the study of \citet{chen20}, every candidate mini-broad absorption line (mini-BAL) system has been checked for the contamination caused by an unrelated intervening absorption line system, such as the damped \lya\ (DLA) or Lyman limit systems (LLSs) formed in intervening galaxies or extended halos \citep[e.g.,][]{Prochaska15, Berg15, Prochaska08}.
We visually inspected both the normalized and un-normalized spectra of every quasar in the catalog, ultimately selecting \target\ for this study due to its potential capability to reveal line-locked systems. 
Our primary objective is to investigate these line-locked systems to better understand the outflow dynamics in quasars. 

Quasar \target\ (also named QSO~2212-179, LBQS~2212-1759) is optically bright ($m_{\rm B} = 18.2$) and has a redshift of $z=2.217$ measured by \citep{Morris91}. 
It was detected in the survey carried out by \citet{Savage79} and  was classified as a broad absorption line (BAL) quasar in the Hewitt and Burbidge catalog \citep{HB87}. The Large Bright QSO Survey \citet[LBQS,][]{Morris91} measured the redshift and confirmed its quasar nature. 
\citet{Korista93} studied the shape of its \civ\ BALs using the Double Spectrograph on the Hale~5m telescope on Mount Palomar.
Quasar~\target\ was found to be X-ray weak after examining the survey data from \textit{ROSAT} \citep{Green95} and the long exposure from \textit{XMM-Newton} \citep{Clavel06}. 
\citet{Gallagher07} examined its mid-infrared and optical properties by constructing a radio through X-ray spectral energy distribution.

\begin{figure*}
\centering
\includegraphics[width=1.0\textwidth]{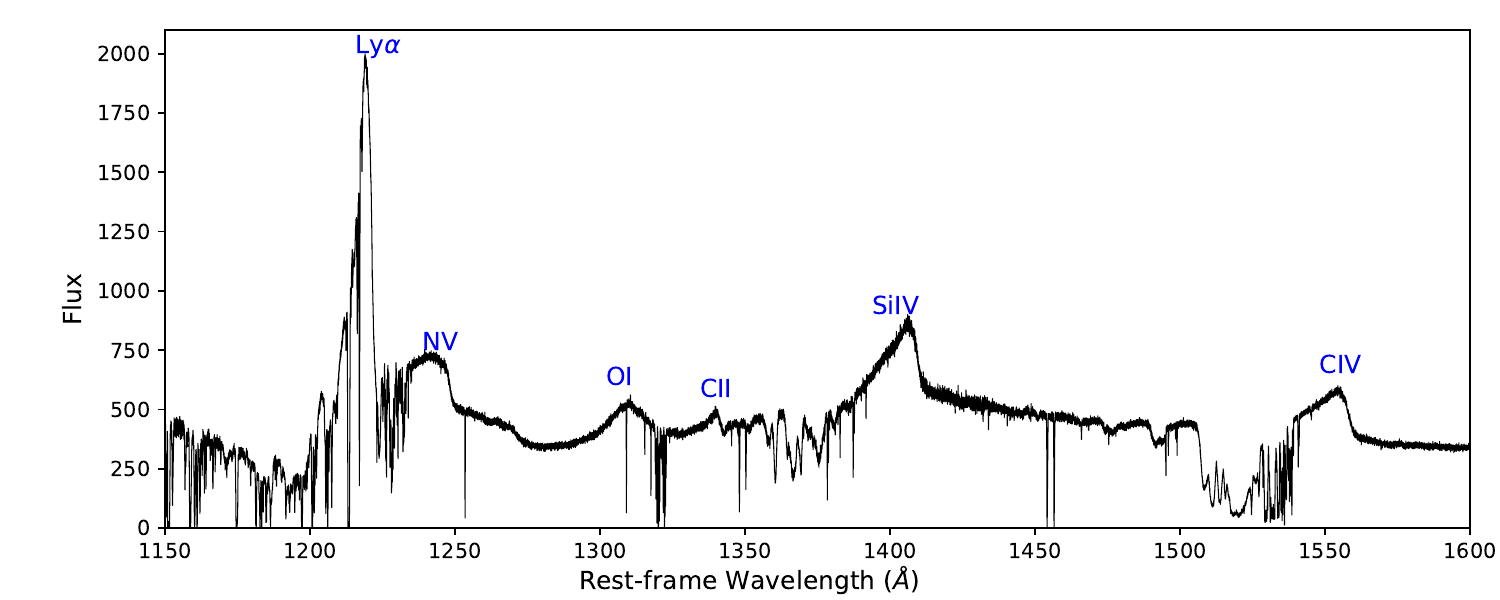}
\caption{UVES spectrum of \target\ at the rest-frame wavelengths (defined by the emission redshift $z=2.217$) showing the absorption lines relative to the broad emission lines. The main emission lines are labeled across the top.\label{fig:spectrum}}
\end{figure*}

UVES is a dual-arm, grating cross-dispersed high-resolution optical echelle spectrograph mounted on the Nasmyth B focus of Unit Telescope 2 of the VLT \citep{Dekker00}. 
The UVES observations of quasar~\target\ (PI: Petitjean; Program ID: 079.A-0251A, 381.A-0634B) were conducted in 2007 and 2008 under a medium seeing of 1.19~arcsec, with a total on-target exposure time of 47784~seconds \citep[see also][]{Murphy19}. 
Fourteen exposures were taken in the UVES standard configuration using the 1'' slit, providing a spectral resolution of $R\sim$40,000 (corresponding to a velocity resolution of $\sim$6.6~\kms\ FWHM) and wavelength coverage of 3283$-$9467~\AA. 
\citet{Murphy19} reduced the data using a modified version of the UVES reduction pipeline based on the optimally extracted method \citep{Horne86, Piskunov02, Zechmeister14, Ma19}. We show the spectrum at the rest-frame wavelengths in \Cref{fig:spectrum}. 
We did not detect time-variable absorption lines from the fourteen exposures taken more than a year apart.
Different observations then were combined to form a single spectrum with a dispersion of 2.5~\kms\ per pixel. The wavelength calibration is done using the ThAr technique \citep{Murphy07}. 
The final continuum-normalized spectrum has a count-to-noise ratio of 21, 52, 51, 67, and 43 per per 2.5~\kms\ pixel at 3500, 4500, 5500, 6500, and 7500~\AA. 

\subsection{Line Component Fits}
To detect possible line-locked signatures in the AAL complexes of \target, we need to identify all possible absorption lines from the UVES spectrum and fit their line profiles. 
Before we identify AALs, we first examine the continuum placement provided in the UVES archive spectra around each line we intend to fit. 
When deemed necessary, we re-fit the continuum using a locally constrained simple power law near the line.
We then identify individual velocity components in the AAL complexes starting with the \civ\ \lam 1548, 1551 doublet.
Finally, we search for other lines at these same redshifts, such as \siiv\ \lam 1394, 1403, \nv\ \lam 1239, 1243, \ovi\ \lam 1032, 1038, and Lyman lines. 
We find AAL complexes of \civ\ \lam 1548, 1551, \siiv\ \lam 1394, 1403, and \nv\ \lam 1239, 1243 doublets from the UVES spectra of \target. 
Unfortunately, we cannot study \ovi\ \lam 1032, 1038 lines because part of the spectrum is missing due to gaps in wavelength coverage between the echelle orders. 
The spectrum may show \lya\ or \siiii\ \lam 1207 lines, but they are singlets and heavily contaminated in the forest, making them difficult to identify. 
Conversely, we can also identifying some \civ\ components using other absorption lines. For example, \civ\ components 4 and 5 in \Cref{fig:civ} are determined through \siiv\ lines. 
All lines identified in the spectra of quasar \target\ are listed in \Cref{tab:J2215}. 

We follow the exact line-fitting procedures described in \citet{Chen18,Chen19} to fit all the lines identified above. 
We limit our fitting to NALs only, excluding the first six \civ\ components due to their association with a much broader \civ\ BAL.
Specifically, we fit every candidate \civ, \siiv, \nv\ NAL using a Gaussian optical depth profile described by:
\begin{equation}
\label{eq:1}
\tau_{v}=\tau_{0}e^{-{(v-v_0)}^2/b^2},
\end{equation}
where $\tau_v$ is the optical depth at velocity $v$, $\tau_{0}$ is the line center optical depth, $v_0$ is the line center velocity, and $b$ is the Doppler parameter. 
The two lines in the doublet are fitted simultaneously with the same velocity shift and $b$ values.
We account for partial covering in our fitting by assuming a spatially uniform light source and homogeneous absorbing medium, with the same optical depth along every sightline.
Thus, the observed intensity at velocity $v$ is given by:
\begin{equation}
\label{eq:2}
\frac{I_{v}}{I_{0}}=1-C_v+C_ve^{-\tau_{v}},
\end{equation}
where $I_{0}$ is the continuum intensity, $I_{v}$ is the measured intensity at velocity $v$, and $C_v$ is the covering fraction of the absorbing medium across the emission source, such that $0 < C_v \leq 1$ \citep{Ganguly99, Hamann97b, Barlow97b}. We also assume that the covering fraction is constant with velocity across the AAL profiles, i.e., $C_v=C_0$, which is a good approximation for the narrow lines in our study.
The fitting of most AALs in the \target\ spectra is generally straightforward. 
However, the fitting procedure needs to be modified when an AAL system is heavily blended with other AAL systems or unrelated lines. 
For AALs blended with unrelated absorption lines at different redshifts, we mask the spectral regions containing these unrelated lines before fitting. 
When two or more doublets are blended together, we fit all these blended doublets simultaneously. These adjustments allows us to obtain consistently reliable fits for all identified AALs.


We then convert the fitted optical depths at the AAL line centroids to the column densities of \civ, \siiv\, and \nv\ using the following relation:
\begin{equation}
\label{eq:3}
\tau_0 = \frac{\sqrt{\pi} e^2}{m_e c} \frac{N_i f_{\rm oscillator} \lambda_0}{b},
\end{equation}
where $N_i$ is the ion column density, $f_{\rm oscillator}$ is the oscillator strength of the corresponding doublet, $\lambda_0$ is the line-center wavelength, and $b$ is the Doppler parameter.
The fitting results and the column densities are presented in \Cref{tab:J2215}.

\section{Results \& Discussion}
\subsection{Line Fitting Parameters}
\begin{figure*}
\centering
\includegraphics[width=0.9\textwidth]{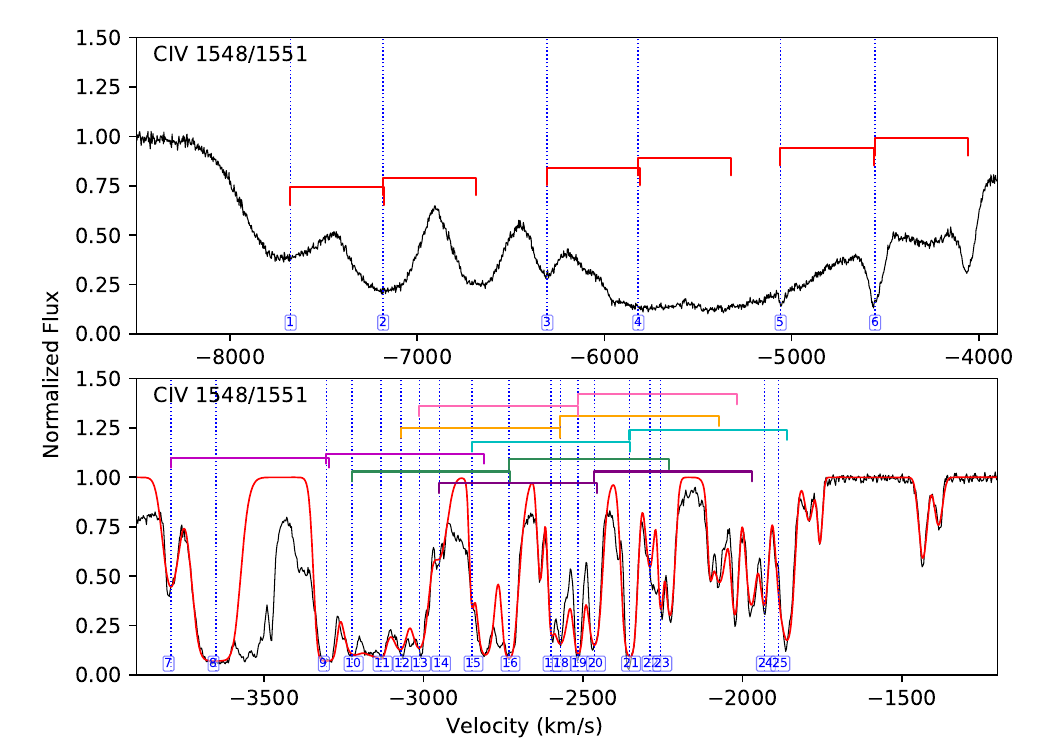}
\caption{Normalized \civ\ line profiles in the VLT-UVES spectra plotted on a velocity scale relative to the quasar redshift $z=2.217$. The spectra are shown in black, and the final fitting lines are shown in red. The blue dash lines are identified components from 1 to 25, and the brackets show the line-locked doublets. The velocities pertain to the short-wavelength lines in the doublets.\label{fig:civ}}
\end{figure*}

\begin{figure*}
\centering
\includegraphics[width=0.9\textwidth]{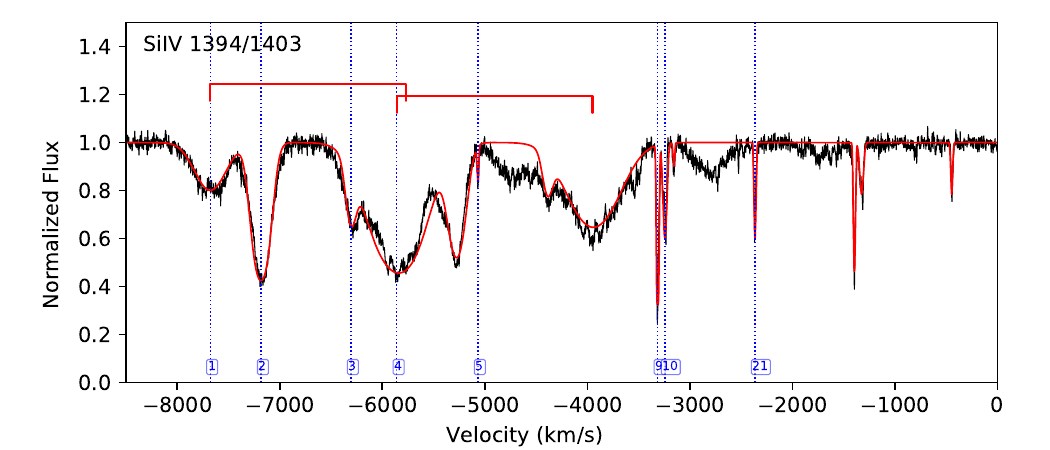}
\caption{Normalized \siiv\ line profiles in the VLT-UVES spectra plotted on a velocity scale relative to the quasar redshift $z=2.217$. The spectra are shown in black, and the final fitting lines are shown in red. The blue dash lines are identified components from 1 to 25, and the brackets show the line-locked doublets. The velocities pertain to the short-wavelength lines in the doublets.\label{fig:siiv}}
\end{figure*}

\begin{figure*}
\centering
\includegraphics[width=0.9\textwidth]{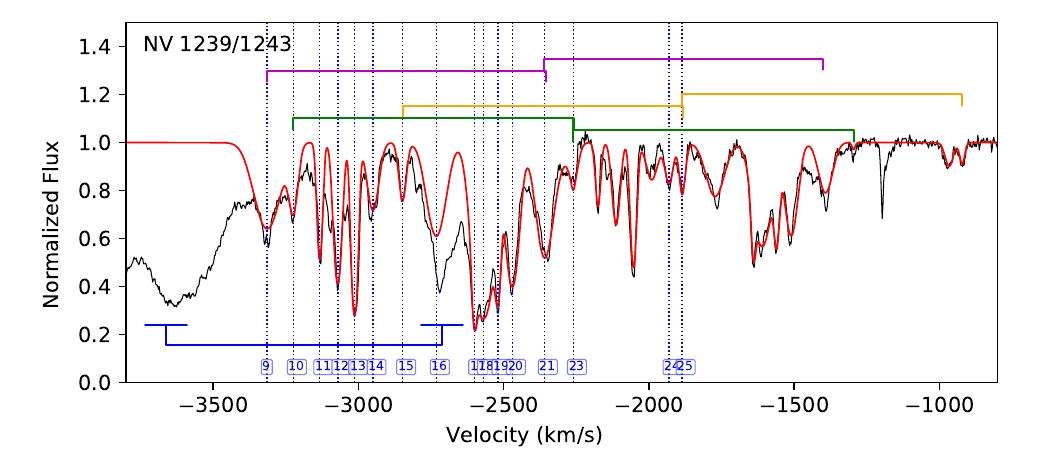}
\caption{Normalized \nv\ line profiles in the VLT-UVES spectra plotted on a velocity scale relative to the quasar redshift $z=2.217$. The spectra are shown in black, and the final fitting lines are shown in red. The blue dash lines are identified components from 1 to 25, and the brackets show the line-locked doublets. The blue bracket in lower left indicates component 16 is blended with unrelated lines. The velocities pertain to the short-wavelength lines in the doublets.\label{fig:nv}}
\end{figure*}

\Cref{tab:J2215} provides the fitted parameters and uncertainties of \civ, \siiv, and \nv\ doublets in the quasar spectra of \target\ from VLT-UVES. 
The table lists the measured quantities, which include the velocity shift, $v$, line identification and rest wavelength, observation wavelength, Doppler parameter $b$, logarithm of column density, and covering fraction $C_0$ for each of the \civ, \siiv, and \nv\ doublet lines separately.
Notes in the last column provide information on blends. The line data are organized in the table according to the redshifts, i.e., component numbers in the first column as depicted in \Cref{fig:civ,fig:siiv,fig:nv}. The uncertainties listed for most of the parameters are 1$\sigma$ errors output from the fitting code, which are due mainly to pixel-to-pixel noise fluctuations in the spectra. For extremely blended lines, we follow the same procedures described in \citet{Chen18, Chen19} to obtain direct estimates or limits on the parameter values.

\Cref{fig:civ} shows our fits to all of the \civ\ AALs included in our final catalog. We derive $v$, $C_0$ and $\log N$ from \Cref{eq:1,eq:2,eq:3} by fitting the \civ\ doublet line profiles. 
Additionally, there are \civ\ mini-BALs in the spectrum, and we have flagged them (component 1 to 6) in the catalog presented in \Cref{tab:J2215}. 
We have not performed line profile fitting on these six \civ\ mini-BALs, as they are part of a blended BAL system.
In \Cref{fig:siiv} and \Cref{fig:nv}, we present our fits to all of the \siiv\ AALs and \nv\ AALs in our final catalog, respectively.

\begin{deluxetable*}{ccccccccccc}
\tablecaption{Individual absorption lines of \target. Columns show component number, absorption redshift ($z_{abs}$) and the corresponding velocity shift ($\textrm{v}$) relative to the emission-line redshift 2.217 \citep{Murphy19}, line identification and rest wavelength, observation wavelength, Doppler b parameter, logarithm of column density, covering fraction, and notes ('bl'=blended with neighboring systems or unrelated lines, if the blended system is also measured, we record bl-X to indicate the component number of the other system, 'bl-BAL'= in the wing of a much broader \civ\ BAL).\label{tab:J2215}}
\tabletypesize{\scriptsize}
\tablehead{
\colhead{$\#$} & \colhead{$z_{abs}$} & \colhead{$\textrm{v}$ (\kms)} & \colhead{ID1} & \colhead{$\lambda_{obs1}$ (\AA)} & \colhead{ID2} & \colhead{$\lambda_{obs2}$ (\AA)} & \colhead{$b$ (\kms)} & \colhead{$\log N$ (\cmN)}  & \colhead{$C_0$} & \colhead{Notes} 
}  
\startdata
1 & 2.1347 & -7678 & \siiv\ 1394 & 4369.14 & \siiv\ 1403 & 4397.26 & $201.5\pm24.5$ & $13.90\pm0.20$ & $0.53\pm0.05$ & bl-4\\ 
 &  &  & \civ\ 1548 & 4853.13 & \civ\ 1551 & 4861.20 & -- & -- & -- & bl-BAL\\
2 & 2.1400 & -7183 & \siiv\ 1394 & 4376.53 & \siiv\ 1403 & 4404.70 & $104.2\pm6.1$ & $14.35\pm0.07$ & $0.61\pm0.02$ & \\ 
 &  &  & \civ\ 1548 & 4861.33 & \civ\ 1551 & 4869.42 & -- & -- & -- & bl-BAL\\
3 & 2.1494  & -6306 & \siiv\ 1394 & 4389.63 & \siiv\ 1403 & 4417.88 & $64.5\pm6.4$ & $13.40\pm0.04$ & $0.66\pm0.07$ & \\ 
 & & & \civ\ 1548 & 4875.89 & \civ\ 1551 & 4884.00 & -- & -- & -- & bl-BAL\\
4 & 2.1546 & -5820 & \siiv\ 1394 & 4396.88 & \siiv\ 1403 & 4425.18 & $320.1\pm8.3$ & $14.73\pm0.25$ & $0.55\pm0.11$ & bl-1 \\ 
 & & & \civ\ 1548 & 4883.94 & \civ\ 1551 & 4892.06 & -- & -- & -- & bl-BAL\\
5 & 2.1627 & -5068 & \siiv\ 1394 & 4408.17 & \siiv\ 1403 & 4436.54 & $10.0\pm2.1$ & $12.18\pm0.17$ & $0.67\pm0.26$ & \\ 
 & & & \civ\ 1548 & 4896.48 & \civ\ 1551 & 4904.62 & -- & -- & -- & bl-BAL\\
6 & 2.1682 & -4550 & \civ\ 1548 & 4904.99 & \civ\ 1551 & 4913.15 & -- & -- & -- & bl-BAL\\ 
7 & 2.1763 & -3799 & \civ\ 1548 & 4917.53 & \civ\ 1551 & 4925.71 & $30.8\pm1.7$ & $\gtrsim13.76$ & $0.85\pm0.15$ & bl-9\\ 
8 & 2.1778 & -3651 & \civ\ 1548 & 4919.85 & \civ\ 1551 & 4928.04 & $51.9\pm3.7$ & $14.93\pm0.07$ & $0.93\pm0.01$ & bl, bl-11 \\ 
9 & 2.1816 & -3314 & \nv\ 1239 & 3941.43 & \nv\ 1243 & 3954.11 & $58.2\pm3.7$ & $14.27\pm0.06$ & $0.62\pm0.05$ & bl-21\\
& & & \siiv\ 1394 & 4434.51 & \siiv\ 1403 & 4463.05 & $11.3\pm0.4$ & $13.35\pm0.06$ & $0.65\pm0.05$ & \\ 
 & & & \civ\ 1548 & 4925.74  & \civ\ 1551 & 4933.93 & $17.1\pm1.5$ & $14.62\pm0.05$ & $0.93\pm0.02$ & bl-7,15\\
10 & 2.1824 & -3244 & \nv\ 1239 & 3942.42 & \nv\ 1243 & 3955.10 & $20.9\pm2.2$ & $13.60\pm0.08$ & $0.57\pm0.12$ & bl-23\\
& & & \siiv\ 1394 & 4435.63 & \siiv\ 1403 & 4464.18 & $24.0\pm1.2$ & $13.10\pm0.04$ & $0.62\pm0.08$ & \\ 
 & & & \civ\ 1548 & 4926.98 & \civ\ 1551 & 4935.17 & $34.4\pm2.2$ & $\gtrsim14.33$ & $0.95\pm0.03$ & bl-16\\
11 & 2.1834 & -3133 & \nv\ 1239 & 3943.66 & \nv\ 1243 & 3956.34 & $12.0\pm0.6$ & $13.49\pm0.02$ & $0.91\pm0.12$ &  \\
 & & & \civ\ 1548 & 4928.52 & \civ\ 1551 & 4936.72 & $11.7\pm0.7$ & $\gtrsim13.57$ & $0.96\pm0.02$ & bl-8,12\\
12 & 2.1841 & -3071 & \nv\ 1239 & 3944.53 & \nv\ 1243 & 3957.21 & $17.7\pm0.8$ & $13.80\pm0.04$ & $0.91\pm0.08$ &  \\
 & & & \civ\ 1548 & 4929.61 & \civ\ 1551 & 4937.81 & $26.5\pm3.9$ & $\gtrsim14.12$ & $0.96\pm0.02$ &bl-11,13,18 \\
13 & 2.1847 & -3013 & \nv\ 1239 & 3945.27 & \nv\ 1243 & 3957.96 & $13.6\pm0.6$ & $13.95\pm0.07$ & $0.89\pm0.06$ &  \\
& & & \civ\ 1548 & 4930.54 & \civ\ 1551 & 4938.74 & $28.9\pm2.4$ & $\gtrsim14.15$ & $0.96\pm0.02$ &bl-12,19 \\
14 & 2.1854 & -2950 & \nv\ 1239 & 3946.14 & \nv\ 1243 & 3958.83 & $26.3\pm2.0$ & $13.56\pm0.05$ & $0.86\pm0.13$ & \\
& & & \civ\ 1548 & 4931.62 & \civ\ 1551 & 4939.82 & $30.1\pm7.3$ & $\gtrsim13.59$ & $0.91\pm0.10$ & bl-20\\
15 & 2.1865 & -2848 & \nv\ 1239 & 3947.50 & \nv\ 1243 & 3960.19 & $17.0\pm1.1$ & $13.70\pm0.16$ & $0.36\pm0.10$ & bl-25\\
& & & \civ\ 1548 & 4933.32 & \civ\ 1551 & 4941.53 & $9.6\pm0.5$ & $\gtrsim14.30$ & $0.73\pm0.25$ & bl-9,21\\
16 & 2.1877 & -2732 & \nv\ 1239 & 3948.99 & \nv\ 1243 & 3961.69 & $41.9\pm2.3$ & $\gtrsim13.87$ & $1.00\pm0.09$ & bl \\
& & & \civ\ 1548 & 4935.18 & \civ\ 1551 & 4943.39 & $17.3\pm0.5$ & $14.04\pm0.05$ & $0.93\pm0.02$ & bl-10\\
17 & 2.1891  & -2600 & \nv\ 1239 & 3950.72 & \nv\ 1243 & 3963.43 & $9.6\pm0.9$ & $13.40\pm0.04$ & $1.00\pm0.10$ &  \\
& & & \civ\ 1548 & 4937.35 & \civ\ 1551 & 4945.56 & $9.9\pm1.6$ & $13.34\pm0.12$ & $1.00\pm0.10$ & \\
18 & 2.1894  & -2570 & \nv\ 1239 & 3951.10 & \nv\ 1243 & 3963.80 & $41.4\pm8.4$ & $14.23\pm0.27$ & $1.00\pm0.05$ &  \\
& & & \civ\ 1548 & 4937.81 & \civ\ 1551 & 4946.03 & $29.2\pm2.9$ & $\gtrsim13.94$ & $1.00\pm0.10$ & bl-12\\
19 & 2.1900 & -2520 & \nv\ 1239 & 3951.84 & \nv\ 1243 & 3964.54 & $12.4\pm0.7$ & $13.56\pm0.03$ & $1.00\pm0.11$ &  \\
& & & \civ\ 1548 & 4938.74 & \civ\ 1551 & 4946.96 & $14.5\pm0.6$ & $\gtrsim13.88$ & $0.98\pm0.02$ & bl-13\\
20 & 2.1906 & -2470 & \nv\ 1239 & 3952.58 & \nv\ 1243 & 3965.29 & $28.4\pm1.0$ & $13.86\pm0.02$ & $0.94\pm0.07$ &  \\
& & & \civ\ 1548 & 4939.67 & \civ\ 1551 & 4947.89 & $21.2\pm0.6$ & $\gtrsim14.01$ & $0.90\pm0.11$ & bl-14\\
21 & 2.1918 & -2365 & \nv\ 1239 & 3954.07 & \nv\ 1243 & 3966.78 & $32.4\pm1.6$ & $14.02\pm0.06$ & $0.54\pm0.07$ & bl-9\\
& & & \siiv\ 1394 & 4448.73 & \siiv\ 1403 & 4477.36 & $12.1\pm0.8$ & $12.81\pm0.04$ & $0.63\pm0.12$ & \\ 
& & & \civ\ 1548 & 4941.53 & \civ\ 1551 & 4949.75 & $21.8\pm0.7$ & $\gtrsim14.18$ & $0.98\pm0.01$ & bl-15,25 \\
22 & 2.1924 & -2290 & \civ\ 1548 & 4942.46 & \civ\ 1551 & 4950.68 & $17.3\pm2.0$ & $\gtrsim13.25$ & $0.95\pm0.03$ & bl\\
23 & 2.1928  & -2260 & \nv\ 1239 & 3955.31 & \nv\ 1243 & 3968.02 & $8.0\pm6.8$ & $12.27\pm0.35$ & $0.34\pm0.30$ & bl-10 \\
& & & \civ\ 1548 & 4943.08 & \civ\ 1551 & 4951.30 & $10.3\pm1.8$ & $\gtrsim13.25$ & $0.94\pm0.03$ & bl\\
24 & 2.1963 & -1931 & \nv\ 1239 & 3959.64 & \nv\ 1243 & 3972.37 & $19.2\pm1.1$ & $13.64\pm0.15$ & $0.35\pm0.10$ &  \\
& & & \civ\ 1548 & 4948.50 & \civ\ 1551 & 4956.73 & $17.1\pm0.6$ & $13.62\pm0.02$ & $0.90\pm0.07$ & \\
25 & 2.1968  & -1886 & \nv\ 1239 & 3960.26 & \nv\ 1243 & 3973.00 & $11.9\pm1.2$ & $13.29\pm0.15$ & $0.28\pm0.19$ & bl-15 \\
& & & \civ\ 1548 & 4949.27 & \civ\ 1551 & 4957.50 & $16.7\pm0.8$ & $\gtrsim13.38$ & $0.92\pm0.03$ & bl-21 \\
\enddata
\end{deluxetable*}

\subsection{Line-locked Signatures}

\begin{figure*}[htbp]
\centering
\includegraphics[width=0.9\textwidth]{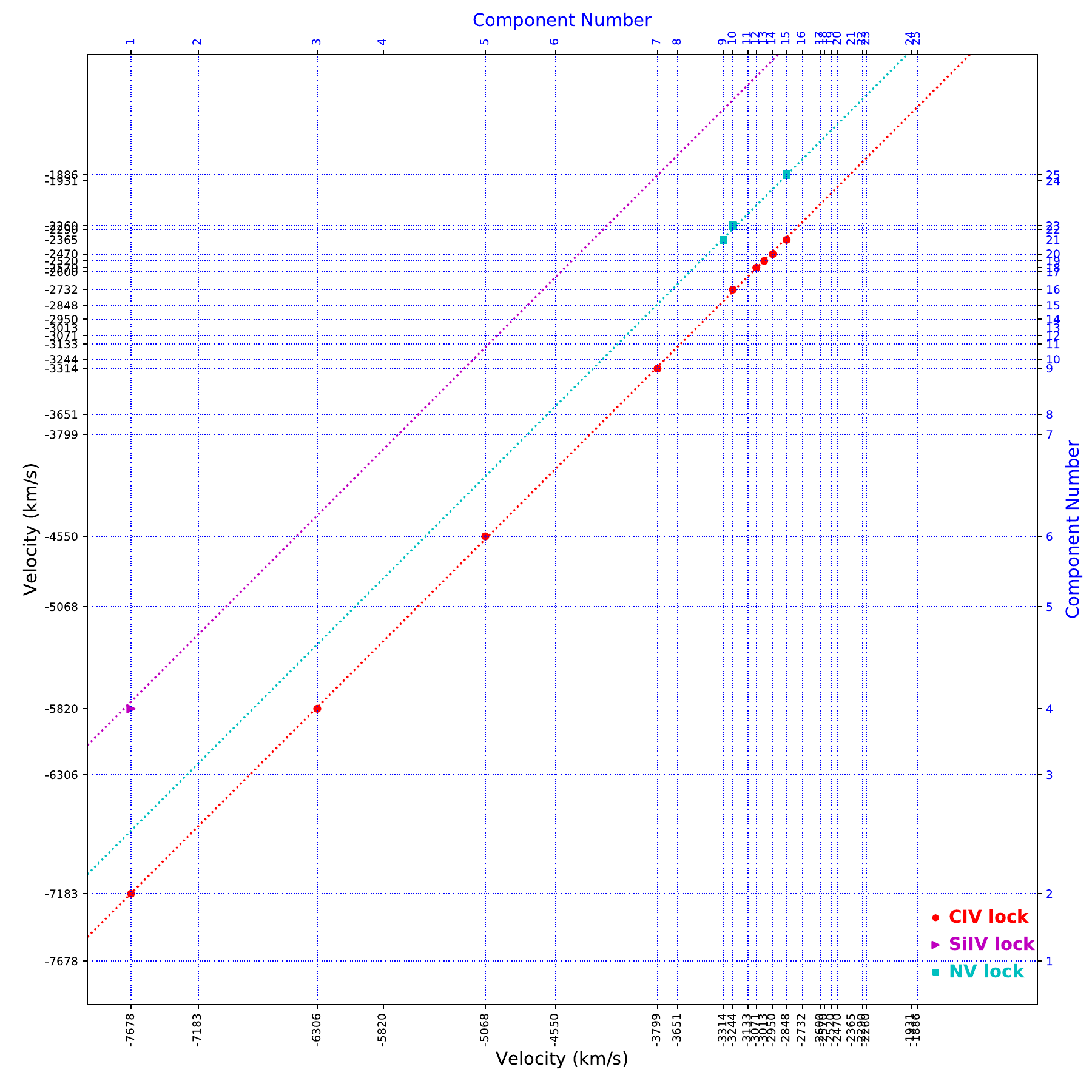}
\caption{The component matrix shows all possible line locks in \civ, \siiv\ and \nv\ based on their velocity separations. We consider two systems are locked when their velocity separation is a close match (with errors less than 10\%) to their laboratory doublet separation (see text for details). All line-locking pairs corresponding to the same ion doublet sit on the same line, where red line for \civ doublet, purple line for \siiv, and green line for \nv. 
\label{fig:matrix}}
\end{figure*}

We search for line-locked pairs corresponding to the \civ, \nv, or \siiv\ doublets in the spectrum of \target\ utilizing the component matrix in velocity space shown in \Cref{fig:matrix}. We skip the usual \ovi\ doublets search because it is not in the wavelength coverage of the \target\ spectra. 
Possible line-locked pairs can be identified in the matrix by drawing a straight line with a slope of $+1$, and an intercept corresponding to the doublet velocity separation, which is $\sim498$~\kms\ for \civ, $\sim 964$~\kms\ for \nv, and $\sim 1939$~\kms\ for \siiv. 
Two absorbers are considered to be line-locked if the velocity difference between them meets the following criterion:
\begin{equation}
    \mid{v_1-v_2}\mid \le min( \rm{b_1}, \rm{b_2}, 0.1*v_{\rm sep}),
\end{equation}
where $v_{\rm sep}$ is the velocity separation of the two lines in the corresponding doublet, and $\rm b_1$ and $\rm b_2$ are the Doppler parameters of the two overlapping lines from the two doublets, respectively. 
The confirmed line-locked pairs are marked using brackets in \Cref{fig:civ,fig:nv,fig:siiv}. 

In the top panel of \Cref{fig:civ}, we identify three line-locked pairs of \civ\ absorbers: pairs (1, 2), (3, 4), and (5, 6), with velocity separations corresponding to the \civ\ $\lambda$1548, 1551 doublet. These lines are mini-BALs blended with a BAL, which made it impossible to accurately fit the column density. 
In the bottom panel of \Cref{fig:civ}, we detect six possible line-locked pairs corresponding to the \civ\ doublets: pairs (7, 9), (10, 16), (12, 18), (13, 19), (14, 20), and (15, 21). At first glance, it appeared that absorbers (7, 9, 15, 21, 25) form the first five successive line-locked \civ\ doublets. However, after carefully checking the fitting, we do not find sufficient evidence to support this claim. 
This underscores the significance of high-resolution spectra in the study of line-locking of quasar outflow.

\begin{figure*}[htb]
\centering
\includegraphics[angle=270, width=0.9\textwidth]{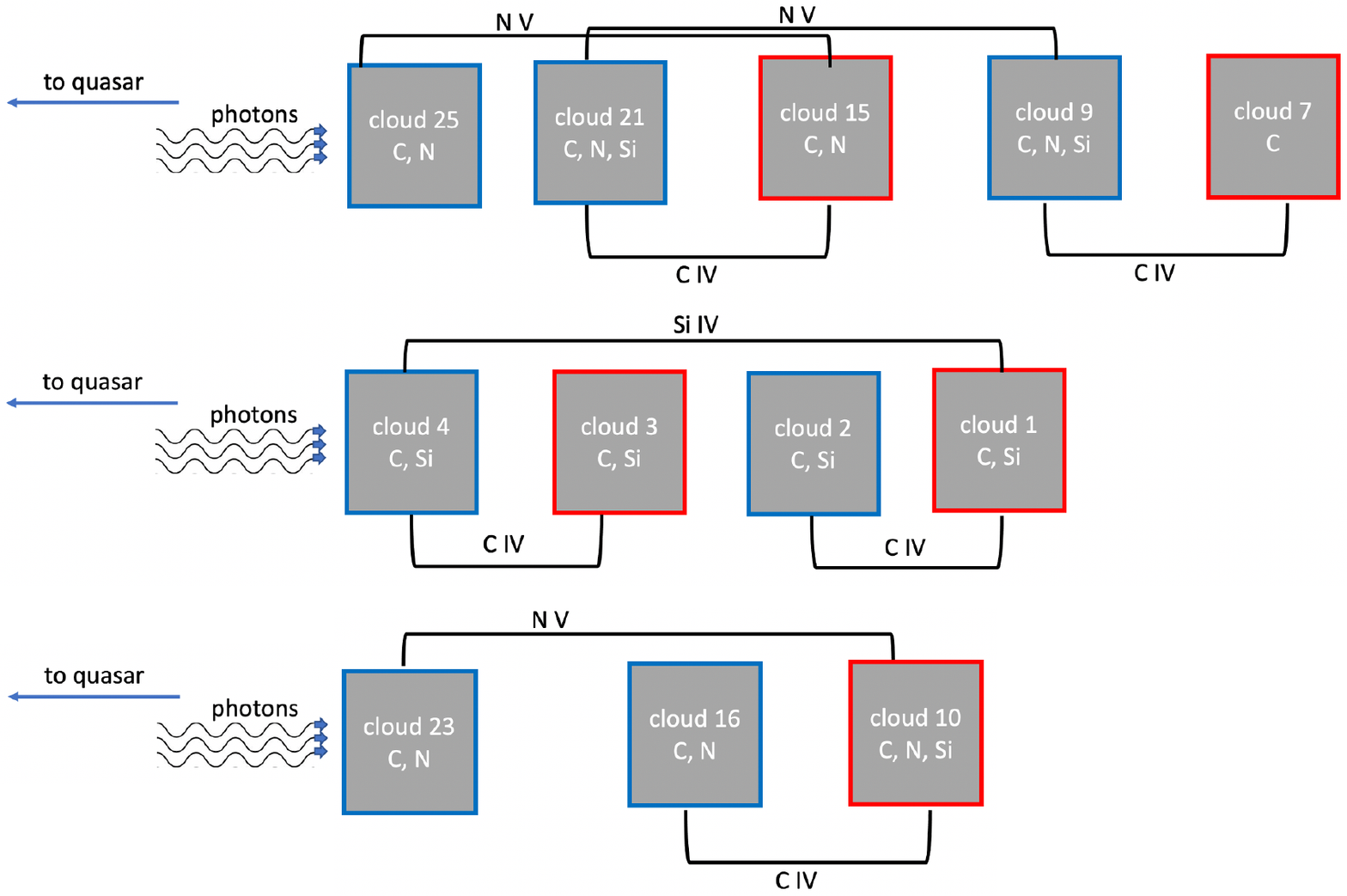}
\caption{Three complex absorber systems line-locked through multi-ion doublets detected in the spectrum of \target. All clouds are exposed to ionizing radiation from the left. We use blue color to mark the shielding clouds in the line-locked pair. The number inside the cloud box is the corresponding component index shown in \Cref{tab:J2215}. We also show if the cloud having \civ, \siiv\ and \nv\ doublets identified.}
\label{fig:multi-lock}
\end{figure*}

We have also detected line-locks in the doublet lines of other ions, including \siiv\ \lam 1393, 1403, and \nv\ \lam 1239, 1243. 
We find three \nv\ NAL pairs, including components (9, 21), (10, 23), and (15, 25), which are marked with brackets in \Cref{fig:nv}.
We identify one \siiv\ NAL pair, comparising absorber components (1, 4), which is marked with red brackets in \Cref{fig:siiv}. 
We find only one \pv\ \lam 1118, 1128 doublet for absorber component 2 and fail to identify any line-locking pairs for \pv\ NALs (see \citet{chen20} for more reference).
It is possible that lines appearing at the doublet separation in observed spectra result from chance alignments of physically unrelated absorption-line clouds. 
However, observations of multiple line-locks in the same spectrum strongly argue for the reality of physical line-locks \citep[e.g.,][]{Ganguly08, Hamann11}. 


From the matrix shown in \Cref{fig:matrix}, we can extract complex line-locking connections involving multiple ionic species. 
We identify three complex systems with three to five absorbers locked together through multi-ionic doublets, illustrated in the cartoon plot in \Cref{fig:multi-lock}. 
In the top panel, we show that components 7, 9, 15, 21, and 25 are locked together through two \civ\ doublets and two \nv\ doublets. In the middle and bottom panels, we show four and three absorbers locked together through \civ\ and \siiv\, \civ\ and \nv\, respectively.
Such alignments shown in \Cref{fig:multi-lock} could be established by chance. 
In order to estimate the probability that such coincidences occur by chance, we randomly populated the observed velocity range with the number of observed components as done by previous studies \citep{Srianand00, Srianand02}. 
We find the probabilities of detecting the three configurations shown from top to bottom in \Cref{fig:multi-lock} are $5\times10^{-4}$, $5\times10^{-4}$, and $4\times10^{-2}$, respectively. 
The small chance alignment probability, together with the presence of multiple line-locked doublets in one quasar, suggests that the line-locking is most probably true \citep{Srianand00, Srianand02, Juranova24}.
How these line-locked components influence each other will be an interesting project for future research.
Historically, theoretical works have often employed a simple two-clouds-one-doublet configuration to investigate the formation of line-locking \citep{Lewis23}. 
Our findings suggest that future theoretical studies of the underlying mechanisms of line-locking should employ more sophisticated models to capture the complexity observed in these systems.

\subsection{Photoionization Model}

While a specific column density of an ion can originate in vastly different physical conditions, constraints from several lines can narrow down the range of possible properties of the absorbing gas. Several of the NALs identified (component 9, 10, 17, 21, 24) have column densities measured from at least two ion species, which are shown in \Cref{tab:J2215}. 
To provide insights into the ionization states of the absorbing systems from the measured ionic column densities above, we used photoionisation models obtained with the \cloudy\ 2023 release \citep{Chatzikos23}.

\begin{figure}[htbp]
\centering
\includegraphics[width=0.5\textwidth]{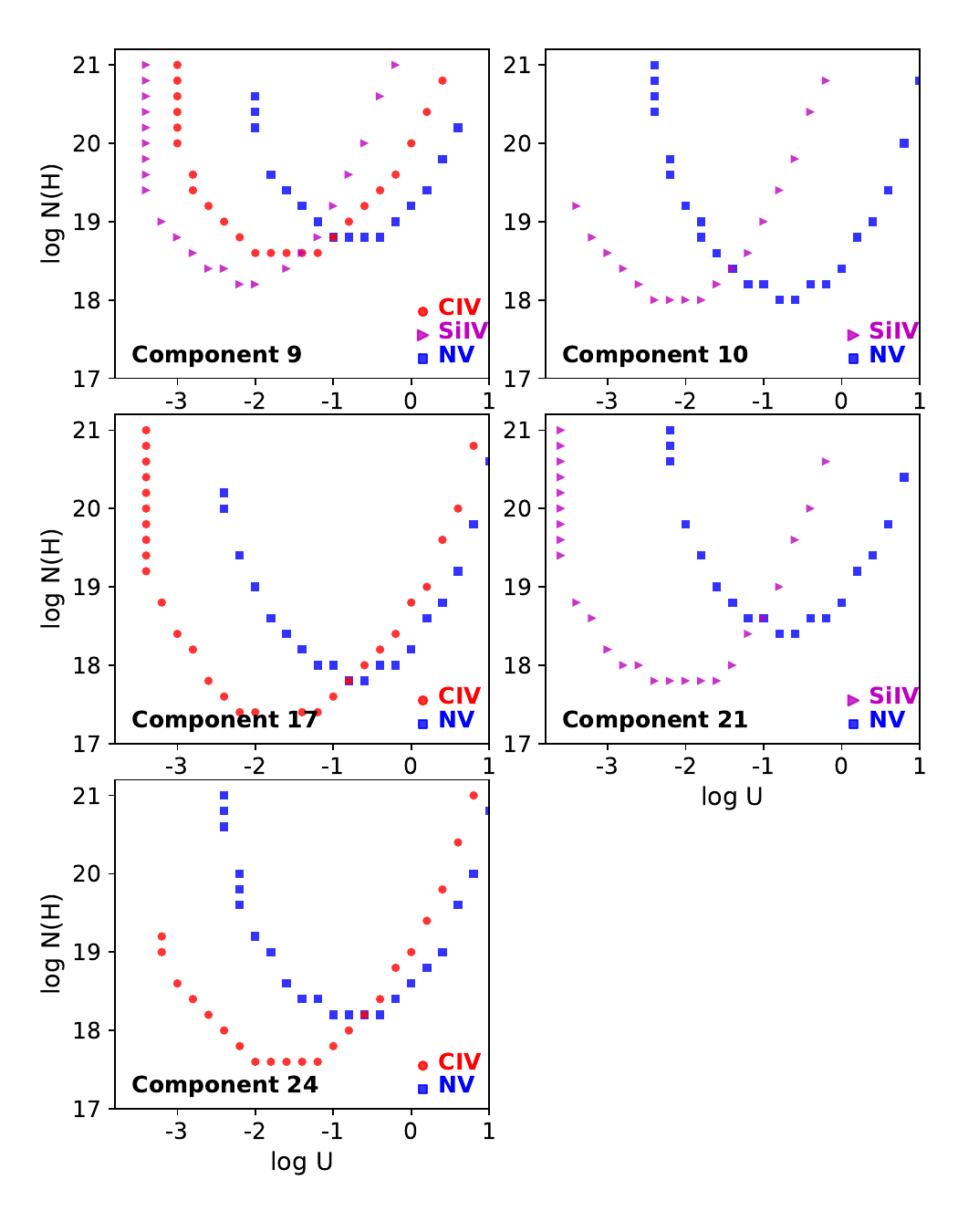}
\caption{Results from the \cloudy\ model for component 9, 10, 17, 21, and 24 at many log $U$-log $N$(H) points assuming solar metallicity. The red, purple and blue symbols mark the results from simulating the column densities of \civ\, \siiv\, and \nv, respectively. The area where different colored symbols meet shows the most likely model parameters.}
\label{fig:cloudy}
\end{figure}

In the \cloudy\ model, we assume the gas is in photoionisation equilibrium with the incident ionising SED, and adopt a solar metallicity.
We set a grid of ionisation parameter $\log U$ and total column density $\log N$(H) with a step of 0.1~dex in each dimension.
The model results from \cloudy\ are shown in \Cref{fig:cloudy}. 
We can derive the hydrogen column density $\log N$(H) and the ionisation parameter $\log U$ for these five absorbers from \Cref{fig:cloudy}. 
The clouds have $\log N$(H) from 17.5 to 19.0 and $\log U$ from 0 to -2.
Among these five absorbers, only components 9 and 21 are line-locked together. 
So we use these two absorbers to test the theory of \citet{Lewis23}. We find these two clouds showing similar values of $\log U$ and $\log N$(H), which seems to support the claim of \citet{Lewis23} that in order to lock together, absorbers need to have similar physical parameters.

By adopting solar metallicity, cloud size scale of 0.1 to 1~pc, and hydrogen column density of $\log N$(H) = 18.5 as derived from \cloudy, we can estimate the masses of a typical absorber to be in the range of $3\times10^{-4} M_\odot$ to $3\times10^{-2} M_\odot$, which seems to be consistent with the Asymptotic Giant Branch (AGB) ejecta model proposed by \citet{Lewis23}. 

Next we estimate the total mass and kinetic energy in the complex line-locked system of \target\ to determine if it can plausibly be important for feedback to the host galaxy’s evolution. 
Following \citet{Hamann11} and \citet{Chen18}, we treat the flow geometry approximately like part of a thin spherical shell, and find the total mass of the complex line-locked system as
\begin{equation}
M \approx 9 \left ( \frac{Q}{20\; \rm{per}\ \rm{cent}} \right)\left(\frac{N_H}{10^{18.5}\; \rm{cm}^{-2}}\right)\left(\frac{R}{10\;\rm{pc}}\right)^2 M_\odot,
\end{equation}
where $N_H = 10^{18.5}$~cm$^{-2}$ is the median column density of the five absorbing systems derived above, $R= 10$ to $100$~pc is a likely radial distance \citep{Hamann11, Lewis23, Hall24}, and Q is the global covering fraction of the outflow, i.e. the fraction of 4$\pi$~steradians covered by the flow as seen from the central continuum source \citep{Hamann00}. 
The value of Q is not constrained by our data, but $Q\sim20$~per~cent is a reasonable guess based on the high detection rates of line-locked \civ\ complex system \citep{Bowler14, Chen21}.
The corresponding kinetic energy, defined by $K=M\textrm{v}^2/2$, is in the range $\sim$$(0.8-80)\times 10^{51}$~ergs if we adopt a typical velocity of $\sim3000$~\kms. 
The characteristic flow time, $\Delta t\sim R/$v, is in the range $\sim(1-10)\times 10^{11}$~s. 
Dividing these energies by the characteristic flow time, yields kinetic energy rates, $\dot{K} \approx (0.8-8)\times 10^{40}$~ergs~s$^{-1}$, that we compare to the near-infrared bolometric luminosity $\sim$$10^{46.8}$ ergs s$^{-1}$ for \target\ \citep{Gallagher07} to derive ratios $\dot{K}/L_{Bol} \approx 10^{-7}-10^{-6}$. 
The upper limit on these ratios are much too small to be important for feedback to the host galaxy evolution, where $\dot{K}/L_{Bol} \gtrsim 0.005$ to 0.05 is believed to be required \citep{Scannapieco04, DiMatteo05, Prochaska09, Hopkins10}. 

\subsection{Origin of Line-locked Systems}
Lines formed in different outflow clumps can become locked at these doublet separations due to shadowing effects in radiatively-driven outflows \citep{Lewis23, Dannen24}. 
Line-locked systems can therefore be a signature of outflows driven by radiative forces in the vicinity of a quasar \citep{Milne26, Scargle73, Braun89, Lin20a, Lin20b}. 
Stable low-velocity and variable high-velocity line-locking systems may favor large-scale and small-scale origins, respectively \citep{Mas-Ribas19a, Mas-Ribas19b}. 
For example, \citet{Lin20a} identified a system with four successive line-locked \civ\ doublets. 
Quasar~\target\ is one of the four quasars known to possess line-locked signatures in \civ, \siiv, and \nv\ at the same time. 
The other three quasars are Q1303+308 \citep{Turnshek84a, Foltz87, Braun89}, J1513+0855 \citep{Srianand02,Chen21}, and SDSS~J092345+512710 \citep{Lin20b}. 
For example, \citet{Lin20b} have studied the \civ\ broad absorption line in SDSS~J092345+512710 and found 11 NALs showing complex line-locking phenomena. 
This suggests also that radiation pressure plays an important role in the dynamics of clumpy outflow clouds in quasar.
The large statistical study by \citet{Bowler14} showed that roughly two-thirds of SDSS quasars with multiple \civ\ NALs at speeds up to $\sim12,000$ km/s have at least one line-locking pair. 
These results indicate that physical line-locking due to radiative forces is both real and common in quasar outflows.

Although the specific physical mechanism is still under debate \citep[see the discussion in][]{Bowler14}, it is widely agreed that line-locked systems are related to the physical process of radiative acceleration \citep{Lewis23}, where successive shielding of the outflowing clouds locks the clouds one by one in outflow velocity \citep{Milne26, Scargle73, Braun89}. 
The case presented here, involving the line-locking of multiple ions, serves as a valuable example for research on the physical mechanism of radiative acceleration.

\begin{figure}[htbp]
\centering
\includegraphics[width=0.48\textwidth]{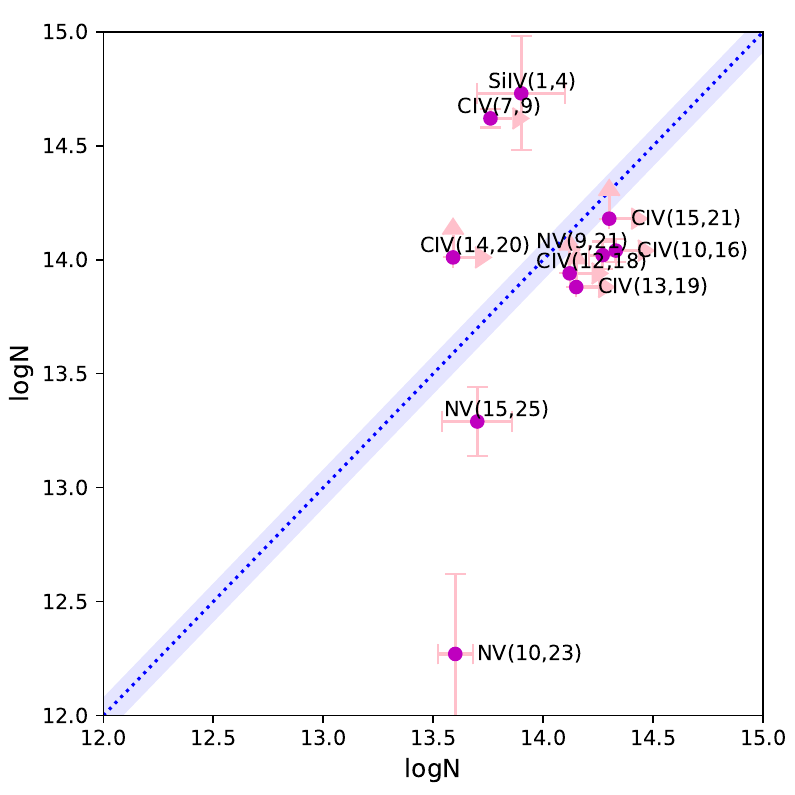}
\caption{Logarithm of column density vs. logarithm of column density of locked systems. The blue dashed line shows the 1:1 ratio. The blue shadow marks the 10\% uncertainties.}
\label{fig:logN}
\end{figure}

\begin{figure}[htbp]
\centering
\includegraphics[width=0.48\textwidth]{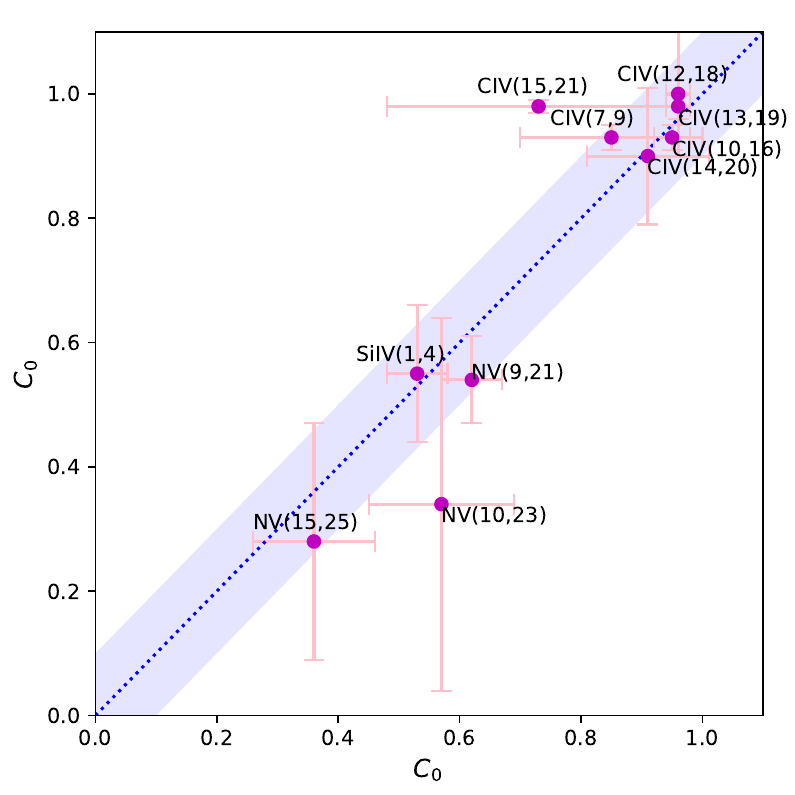}
\caption{Covering fraction $C_0$ vs. covering fraction $C_0$ of locked systems. The blue dashed line shows the 1:1 ratio. The blue shadow marks the 10\% uncertainties.}
\label{fig:C0}
\end{figure}

Using simulations, \citet{Lewis23} suggests the clouds' properties need to be fine-tuned to within 1\% for the clouds to be line-locked. They argue that the prevalence of line-locking in quasar spectra supports the idea that the ejected envelopes of AGB stars are one of the possible origins of the outflowing clouds observed as narrow absorption-line systems.
Our findings can be used to test this scenario.
We have identified a total of 25 absorption-line systems, including 6 mini-BALs and 19 NALs.
All 6 mini-BALs are line-locked, and 12 out of 19 NALs are line-locked, which represents a notably high percentage. This suggests that line-locking should be easily formed.
We also compare the derived ion column densities and covering fractions of the line-locked pairs in \Cref{fig:logN} and \Cref{fig:C0}.
We find that the derived column densities are quite similar for most of the the \civ\ pairs, though this is not the case for every pair.
The covering fractions of the locked pairs appear consistent with each other, as shown in \Cref{fig:C0}, where most of the data points fall near the 1:1 ratio line. 
This generally supports the theoretical study by \citet{Lewis23}, which argues that two absorbers need to have similar physical properties to be locked together.

However, we also find evidence that several clouds can be locked together through different-ion doublets. For example, as shown in \Cref{fig:multi-lock}, there are three complex AALs with three to five absorbers locked together through multi-ionic doublets. 
The existence of these complex line-locking phenomena indicate that it is not enough to study the line-locking using a simple model of two clouds and one ionic doublet, like the one from \citet{Lewis23}. 
And this is the reason that why in reality we may not need the 1\% fine-tuning derived by \citet{Lewis23}, otherwise we should not see so many line-locking signatures as suggested by \citet{Bowler14} and \citet{Chen21}. 
Further theoretical modeling and detailed simulations are needed to reveal the mechanisms behind these line-locking phenomena.

\section{Conclusion}

We present a detailed investigation of the line-locking signatures shown in QSO~\target\ using the high-resolution and high-signal-to-noise archived VLT-UVES spectra. 
Our study is much more sensitive to weak and narrow AALs than all previous outflow line-locking surveys using medium-resolution spectra such as from the SDSS. 
A total of 25 associated absorption line systems were identified, including 6 mini-BALs and 19 NALs. 
We then fit every \civ, \nv, and \siiv\ NALs to obtain basic line properties like velocity shift, column density and the line-of-sight covering fraction, $C_0$. 
We find all 6 mini-BALs are line-locked, and 12 out of the 19 NALs are line-locked, representing a notably high percentage.
The covering fractions derived are varying from $\sim$0.4 to $\sim$1, which supports that a large covering fraction is needed to produce line-locked systems. 
The presented case study of \target\ indicates the high degree of clumpiness of the outflowing materials from the quasar, and the significance of the radiative forces in driving these outflowing materials. 

We also find several clouds can be locked together through multi-doublets, which indicates it is not enough to study the line-locking phenomena using a simple model of 2 clouds and 1 ionic doublet, like the one studied in \citet{Lewis23}. 
And this is the possible reason that why we do not need the 1\% fine-tuning as required by \citet{Lewis23}. 
Further theoretical modeling and simulation are needed to reveal the mechanisms behind line-locking phenomena.
Our findings suggest that high-resolution, high-SNR spectra of quasars are crucial for studying the line-locking mysteries and can increase our understanding of the quasar outflow mechanism.

\section*{acknowledgments}

We thank the anonymous referee for helpful comments and suggestions. 
We acknowledge Prof. Michael Murphy for the quasar spectrum. This work was supported by the National Natural Science Foundation of China (12103097, 12073092). CC acknowledges the support of the "Three Levels" Talent Construction Project of Zhuhai College of Science and Technology. Zhicheng He acknowledges the support of the National Natural Science Foundation of China (nos.12222304, 12192220, and 12192221).

\bibliography{reference}
\bibliographystyle{aasjournal}

\end{document}